\documentclass[sigconf,natbib=false,anonymous=false]{acmart}

\copyrightyear{2026}
\acmYear{2026}
\setcopyright{cc}
\setcctype{by}
\acmConference[DAC '26]{63rd ACM/IEEE Design Automation Conference}{July 26--29, 2026}{Long Beach, CA, USA}
\acmBooktitle{63rd ACM/IEEE Design Automation Conference (DAC '26), July 26--29, 2026, Long Beach, CA, USA}
\acmDOI{10.1145/3770743.3804000}
\acmISBN{979-8-4007-2254-7/2026/07}

\RequirePackage[
  datamodel=acmdatamodel,
  style=acmnumeric,
  ]{biblatex}
\usepackage[nolist]{acronym}
\usepackage{algorithmic}
\usepackage{graphicx}
\usepackage{enumerate}
\usepackage{enumitem}
\usepackage{circledsteps}
\usepackage{textcomp}
\usepackage{listings}
\usepackage{stfloats}  
\usepackage{colortbl}
\usepackage{multirow}
\usepackage{subcaption}
\begin{acronym}
    \acro{afl}[AFL]{American Fuzzy Lop}
    \acro{soc}[SoC]{System-On-Chip}
    \acro{put}[PUT]{Program Under Test}
    \acro{mmio}[MMIO]{Memory-Mapped Input/Output}
    \acro{vcml}[VCML]{Virtual Components Modelling Library}
    \acro{iss}[ISS]{Instruction Set Simulator}
    \acro{fss}[FSS]{Full-System Simulator}
    \acro{vp}[VP]{Virtual Prototype}
    \acro{tlm}[TLM]{Transaction-Level Modeling}
    \acro{dbt}[DBT]{Dynamic Binary Translation}
    \acro{avp}[AVP]{Arm Virtual Platform}
    \acro{nvic}[NVIC]{Nested Vector Interrupt Controller}
    \acro{ftl}[FTL]{Fast Translation Library}
    \acro{os}[OS]{Operating System}
    \acro{mcu}[MCU]{Microcontroller}
    \acro{adc}[ADC]{Analog-to-Digital Converter}
    \acro{rtc}[RTC]{Real-Time Clock}
    \acro{hal}[HAL]{Hardware Abstraction Layer}
    \acro{isr}[ISR]{Interrupt Service Routine}
    \acro{ecu}[ECU]{Electronic Control Unit}
    \acro{rtos}[RTOS]{Real-Time Operating System}
\end{acronym}
\usepackage{tikz} 
\usepackage{pgfplots} 
\usepackage{xcolor} 
\definecolor{rwth_blue}{RGB}{0,84,159}
\usepackage{tcolorbox} 
\tcbuselibrary{listings}
\definecolor{mplblue}{HTML}{1f77b4}
\definecolor{mplorange}{HTML}{ff7f0e}
\definecolor{mplgreen}{HTML}{2ca02c}
\usetikzlibrary{patterns,patterns.meta}

\definecolor{dkgreen}{rgb}{0,0.6,0}
\definecolor{gray}{rgb}{0.5,0.5,0.5}
\definecolor{mauve}{rgb}{0.58,0,0.82}

\newcommand\myCircled[2][]{\ifmmode
\Circled[fill color=white,inner color=black,#1]{\mathsf{#2}}
\else
\Circled[fill color=white,inner color=black,#1]{\sffamily#2}
\fi
}

\begin{document}


\title[Stateful Embedded Fuzzing Using Peripheral-Accurate SystemC Virtual Prototypes]{Stateful Embedded Fuzzing with\\Peripheral-Accurate SystemC Virtual Prototypes}

\author{Chiara Ghinami, Igor Pontes Tresolavy, Luis Seibt, Nils Bosbach, Rainer Leupers}
\email{{ghinami,pontes,seibt,bosbach,leupers}@ice.rwth-aachen.de}
\affiliation{%
  \institution{RWTH Aachen University}
  \city{Aachen}
  \country{Germany}
}

\renewcommand{\shortauthors}{Ghinami et al.}

\begin{abstract}
The increasing complexity of embedded software has made comprehensive manual testing impractical, motivating the use of automated techniques such as fuzzing. Coverage-guided fuzzers like AFL++ have shown strong results for conventional software but remain challenging to apply effectively in embedded contexts, where peripheral behaviors play critical roles. Existing approaches either use fast user-mode simulators, sacrificing peripheral realism, or rely on full-system simulators with manual instrumentation, limiting applicability to large-scale software.

In this work, we present a novel framework that integrates AFL++ with a stateful SystemC-TLM virtual prototype to enable realistic fuzzing of embedded software. Fuzzer-generated inputs are injected directly into peripheral models, allowing peripherals to trigger natural side effects such as interrupts and FIFO updates.

By integrating fuzzing with full-system simulation, our framework advances the effectiveness of pre-silicon testing for embedded systems. Results on embedded workloads show that our approach eliminates false positives while maintaining comparable code coverage and execution performance as state-of-the-art tools.
\end{abstract}




\maketitle

\section{Introduction}
In the past decade, the complexity of embedded software has increased significantly to cope with real-world problems, making manual software testing practically impossible. Fuzzing, a popular testing technique~\cite{10.1145/3623375,10.1145/3512345}, feeds random inputs to the \ac{put} and, being fast and lightweight, avoids issues, such as the state explosion, that affect symbolic execution~\cite{10.1145/3623375}.

\begin{figure}[t!]
\centerline{\includegraphics[totalheight=3.5cm]{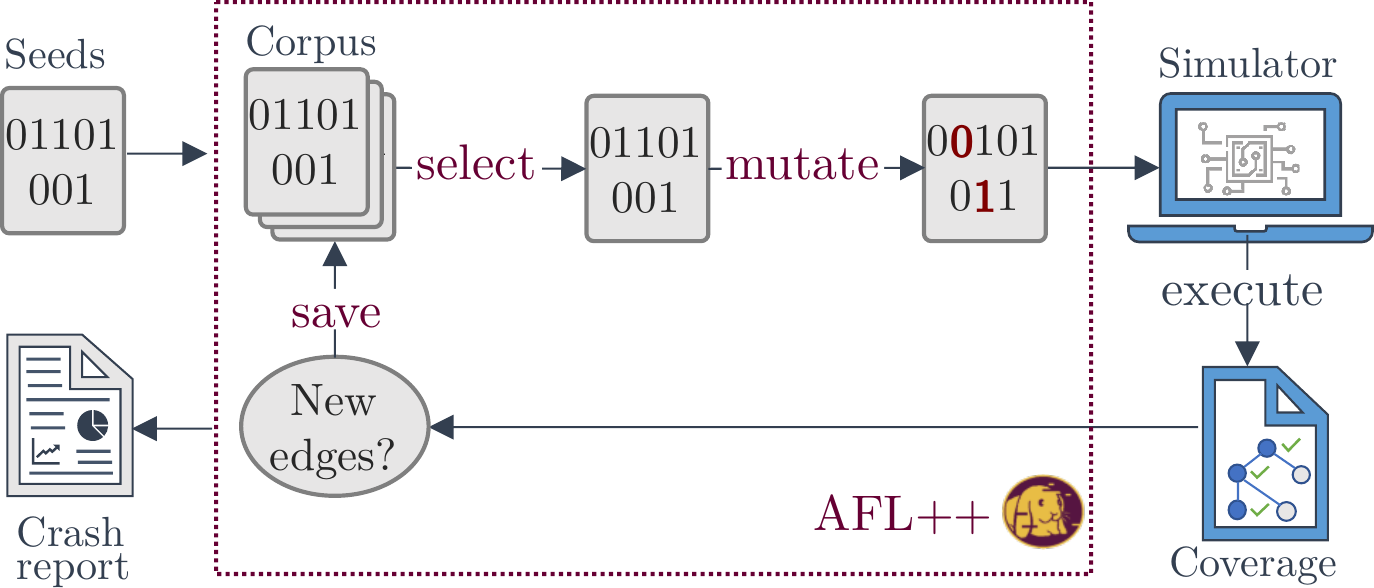}}
    \caption{Simulation-based fuzz testing.}
    \label{fig:framework}
\end{figure} 

AFL++~\cite{fioraldi2020afl++}, a community-maintained fork of the AFL fuzzer~\cite{AFL}, is widely used in industry and research for coverage-guided testing. Recently, significant research has focused on enabling embedded software fuzzing by integrating AFL++ with simulators. As shown in Fig.~\ref{fig:framework}, the fuzzer takes a seed from the user, mutates it, and gives it as input to the \ac{put} that runs on the simulator. During the execution, the simulator gathers code coverage information. Based on that, the fuzzer chooses to discard the input or further use it and add it to the corpus. 
Functional simulators such as QEMU~\cite{10.5555/1247360.1247401} offer fast software execution along with debugging capabilities, making them ideal for early-stage software testing and validation.

For this purpose, both user-mode and full-system simulators are commonly used. User-mode simulators do not simulate the target \ac{os} and peripherals, which limits them to user-space applications. Instead, they forward the system calls of the target software to the host \ac{os}.
In contrast, full-system simulators, or \acp{vp}, model the entire hardware platform, including peripherals, enabling the simulation of the complete target software stack. They provide higher execution fidelity at the cost of reduced performance.

Many research efforts favor user-mode simulators for embedded fuzzing~\cite{10.5555/3489212.3489280,feng2020p2im,scharnowski2022fuzzware,10.5555/3361338.3361415}. These approaches learn which register values influence the control flow and inject them accordingly. While effective in increasing execution speed, frameworks such as Fuzzware~\cite{scharnowski2022fuzzware} and P2IM~\cite{feng2020p2im} fail to model peripheral causality (interrupts, FIFO and register updates), which can result in unrealistic execution traces and high false positives count. 
Other works have employed full-system simulators for fuzzing~\cite{10.1145/3386263.3406899,9218694}. However, these approaches require the modification of peripheral models or manual \ac{put} instrumentation, limiting their broader adoption.

Unlike prior works that rely on program analysis to infer register values, our framework integrates AFL++ with a full-system SystemC-TLM \ac{vp} and introduces a novel injection mechanism that (1) intercepts \ac{mmio} accesses and (2) injects fuzzer-generated values directly into virtual peripheral models on selected \ac{mmio} events, allowing peripherals to produce realistic side effects, without any modifications required to either the peripheral models or the fuzzed source code.
Motivated by recent demonstrations of attacks that inject malicious data through communication interfaces (e.g., CAN and UART)~\cite{car_theft,uart_attack,can_attack}, we target software that interacts directly with \ac{soc} peripherals, such as bare-metal applications, \ac{rtos} device drivers, and \acp{hal} to test whether external data can trigger software faults. 

In several experiments, we show that our framework outperforms prior work in execution fidelity, when evaluated on embedded workloads that exercise a wide range of peripherals.
With this work, we bridge the gap between fuzz testing and full-system simulation, enabling more realistic and effective testing of embedded systems.

\section{Background}
In this section, we provide an overview of \ac{vp} and fuzzing, followed by a summary of existing approaches that leverage simulators for embedded fuzzing.
\subsection{Virtual Prototypes}
A \ac{vp} is a software model of a hardware system that executes the same software as the physical hardware, producing functionally equivalent outputs. To showcase our approach, we utilize SIM-A, a proprietary ARM-based VP developed by MachineWare~\cite{Machineware}. SIM-A is a \textit{functional} simulator, meaning it focuses on the logical correctness of computation.
Unlike \textit{cycle-accurate} simulators, which capture microarchitectural details, functional simulators assume a fixed number of cycles per instruction, ignoring pipeline stages, stalls, or cache effects, to maximize the execution speed.

SIM-A is implemented using SystemC-~\ac{tlm}~\cite{systemc2025}, a C++-based modeling framework that abstracts low-level hardware behavior and provides better modularity and customizability compared to traditional C-based functional simulators such as QEMU~\cite{10.5555/1247360.1247401}. Thanks to the SystemC-TLM standardized interface, that provides a common way to connect components, our approach is not tied to SIM-A specifically and can be applied to any SystemC-TLM-based VP with minimal adaptation. For peripheral models, we rely on the open-source Virtual Components Modeling Library (VCML)~\cite{vcml}. VCML provides a rich set of pre-defined peripheral components that can be directly connected to a SystemC CPU model, enabling the simulation of a broad range of SoC configurations. Leveraging these existing peripheral models, originally developed for general VP use cases, allowed us to significantly reduce the implementation effort while ensuring realistic device behavior.


\subsection{Fuzzing}
Fuzzing is a software-testing technique that aims to uncover bugs in the~\ac{put} by providing unexpected or malformed inputs, often triggering corner-case behaviors.

The most adopted taxonomy classifies fuzzers based on how much information is used from the program execution~\cite{manes2019art}. \textit{Black-box} fuzzers operate without any feedback from program execution. At the opposite end, \textit{white-box} fuzzers select the input based on extensive program analysis, such as symbolic execution~\cite{godefroid2008automated}. In between, \textit{grey-box} fuzzers use lightweight execution feedback to guide input generation, improving coverage and bug detection over black-box approaches, while avoiding the high overhead of white-box analysis~\cite{10.1145/3611019,chen2018angora,fioraldi2020afl++,fioraldi2022libafl}. This makes grey-box fuzzers well-suited for large-scale and complex applications. 
Among grey-box fuzzers, AFL++ is a widely used mutation-based fuzzer that relies on code coverage feedback and a genetic algorithm to generate inputs that explore new execution paths in the \ac{put}~\cite{fioraldi2020afl++}.

When fuzzing native desktop applications, the program is typically instrumented to expose runtime feedback (e.g., code coverage) to the fuzzer. Instead, when fuzzing with a simulator, it is usually the simulator itself that collects coverage information, e.g., by tracking at runtime which basic blocks are translated or executed~\cite{fioraldi2020afl++}.

\subsection{Related Work} \label{related_work}
In recent years, other studies have focused on fuzzing embedded system software using simulators. These techniques can be broadly categorized based on the type of approach they employ, which we refer to as \textit{stateless} and \textit{stateful}.

\subsubsection*{Stateless VP-based fuzzers}
This category includes works that utilize user-mode simulators to perform embedded fuzzing. Because these simulators do not model peripherals, every \ac{mmio} read is treated as a program input that the fuzzer must provide. We refer to them as \textit{stateless} because the peripheral state is not preserved within the program execution. For example, consecutive reads from the same register may return unrelated values, as no device logic maintains internal consistency. HALucinator~\cite{10.5555/3489212.3489280} replaces the functions of the \ac{hal} in microcontroller firmwares and redirects peripheral accesses to a custom peripheral that supplies fuzz data. This approach requires manual effort and a deep understanding of the target firmware.
FirmAFL~\cite{10.5555/3361338.3361415} uses a full-system simulator during the OS boot phase, but then switches to user-mode execution. When the firmware requests data from a peripheral, their framework supplies a fuzzed value. P2IM~\cite{feng2020p2im} and Fuzzware~\cite{scharnowski2022fuzzware} build on this idea. They introduce respectively an explorative firmware technique and a symbolic execution engine to deduct the firmware logic and restrict the state space of the fuzzer-generated value. In Section~\ref{comparison}, we show that these techniques maximize the execution speed but tend to produce a high number of false positives.

\subsubsection*{Stateful VP-based fuzzers}
This category includes works that integrate a full-system simulator with the fuzzer, and closely relate to our work. Both Herdt et al.~\cite{10.1145/3386263.3406899} and Yang et al.~\cite{9218694} connect a coverage-guided fuzzer to a SystemC-TLM \ac{vp}. However, their injection mechanisms, which allow peripherals to receive fuzzer-generated inputs, require manual firmware modifications. For instance, Herdt et al.~\cite{10.1145/3386263.3406899} introduce a dedicated fuzzing peripheral requiring the software to read from it to consume input data. These approaches limit scalability for complex firmware and is impractical for closed-source software.

In contrast to previous stateful frameworks, we propose generic injectors that periodically send fuzz data to the peripherals. Our approach requires no firmware modifications and is therefore suitable for testing closed-source programs.

%
%
%

\section{Fuzzing Framework}
In this section, we present our framework and the adaptations required in the \ac{vp} so that it can efficiently work with the fuzzer. We then describe how the framework employs injector modules to send fuzz data to peripheral models.

\begin{figure}[t!]
\centerline{\includegraphics[totalheight=5cm]{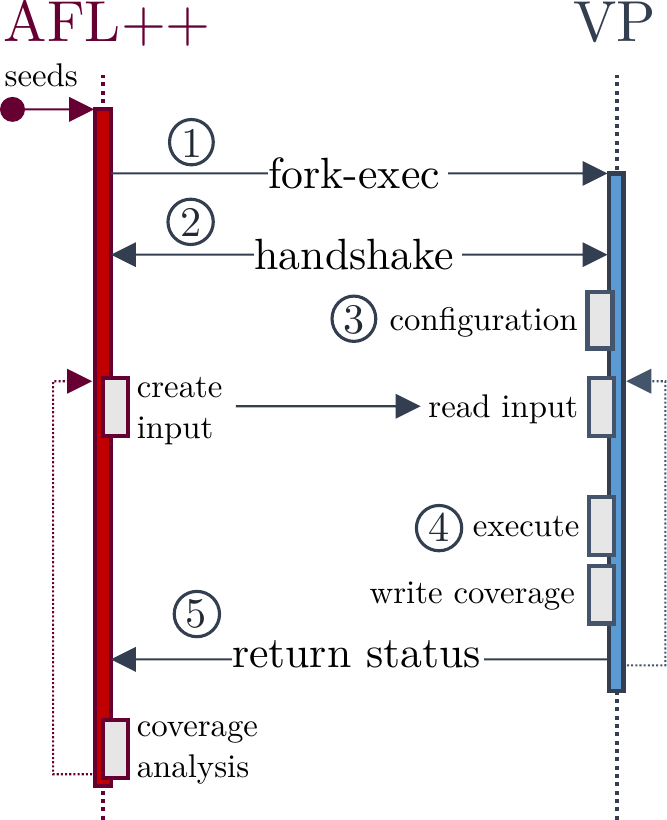}}
    \caption{The steps of the fuzzing workflow.}
    \label{fig:workflow}
\end{figure}

\subsection{Workflow} \label{design}
Because AFL++ already supports QEMU-based fuzzing, no changes to the fuzzer were needed to add a new \ac{vp}. In contrast, SIM-A was enhanced to gather code coverage, detect errors and communicate with the fuzzer.
In Fig.~\ref{fig:workflow}, we show the steps of VP-based fuzzing: 
\begin{itemize}
    \item[\myCircled{1}] The AFL++ fuzzer forks and spawns the \ac{vp} process.
    \item[\myCircled{2}] A handshake procedure begins, during which the fuzzer verifies that the new process is up and running. The \ac{vp} and the fuzzer communicate via pipes and shared memory regions. During the handshake, the fuzzer creates and configures two shared memories: one for fuzzing inputs and one for code-coverage data.
    \item[\myCircled{3}] The \ac{vp} is configured. In this phase, e.g., the exit and error symbols or addresses are set, so that the VP can detect exit conditions and crashes. 
    \item[\myCircled{4}] After the handshake completes, the \ac{vp} starts the execution. While running, injectors periodically read data from the shared memory and send them to the peripherals (see Section~\ref{injectors} for details). During execution, the \ac{vp} records which basic blocks are being translated for code coverage purposes.
    \item[\myCircled{5}] When execution finishes, the \ac{vp} informs the fuzzer whether an error occurred and returns the collected code coverage.
\end{itemize}
The \ac{put} is executed in a loop, once step \myCircled{5} is over, the \ac{put} is restarted with a new input.

\subsection{Input Injection} \label{injectors}
An injector is a module connected to a peripheral via a TLM socket and to the fuzzer via shared memory. In our setup, the \ac{put} usually acts as the communication initiator, issuing reads and writes, or configuring frames. Injectors therefore emulate the counterpart (target/peer) and supply fuzzer inputs to the peripheral model. 

\subsubsection*{Design principles}
The injector design addresses four key aspects:
\begin{enumerate}
    \item Support different communication styles (initiator–target like I\textsuperscript{2}C and peer-to–peer like UART/CAN).
    \item Minimize implementation effort maintaining generality.
    \item Determine at what point of the firmware execution to start the injection of data.
    \item Support different modes of interaction between the CPU and the peripheral (i.e., polling and interrupt).
\end{enumerate}

For peer-to-peer protocols, the injector will generally send data periodically. For initiator–target protocols, it responds to peripheral requests. To implement periodic injection, we spawn an \texttt{SC\_THREAD} inside the injector that periodically transmits data. 

\begin{figure}[t!]
\centerline{\includegraphics[totalheight=5cm]{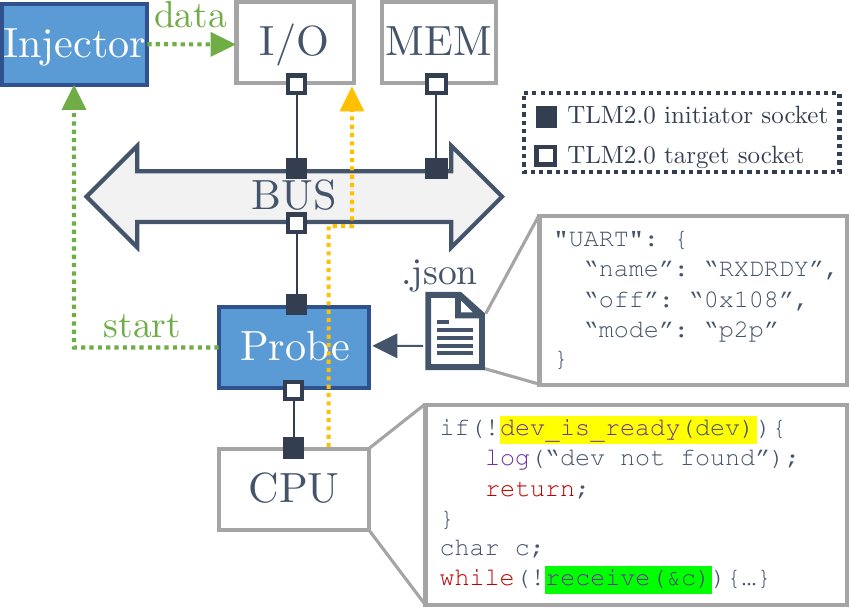}}
    \caption{Injection for peer-to-peer communication.}
    \label{fig:injection}
\end{figure}

Communication protocols commonly follow different conventions. For example, the UART receives one byte at a time while the CAN works with frames. The majority of the injector code, which involves, e.g., shared memory handling, is identical across injectors. Only the routine that packages and emits data according to the peripheral protocol differs. This minimizes duplication while making it simple to add support for a new peripheral, which only needs implementing the protocol-specific packing/send function and registering the injector.
We adopt a \textit{one injector per peripheral} design: each peripheral instance has a dedicated injector. This choice supports different communication logics, while keeping the implementation effort low.

\subsubsection*{Trigger detection and configuration}
To avoid sending fuzz data before a peripheral is configured by the firmware, which would result in the data being discarded by the peripheral model, the injector thread starts after a trigger event is observed (e.g., the enabling of peripheral interrupts, or a specific register access used for polling).
A lightweight \textit{probe} component sits between the CPU and the bus (Fig.~\ref{fig:injection}) and observes register accesses. The probe detects trigger conditions, identifies the affected peripheral instance, and instructs the corresponding injector to spawn the \texttt{SC\_THREAD}. Injector trigger conditions and per-peripheral options are configurable in a JSON file that the VP reads during initialization (Step \myCircled{3} in Section~\ref{design}). This avoids hardcoding injection rules into the code and allows for easy and adaptable per-peripheral customization. In Fig.~\ref{fig:injection}, we show the injection mechanism in the case of peer-to-peer communications. The code highlighted in yellow represents an access to the peripheral that does not trigger the injector. In green, we show a function that triggers the injector by accessing one of the registers specified in the JSON file.
For both polling and interrupt-based communications, the probe watches accesses to a configurable register (settable in the JSON) and starts injection on the first access. This trigger event could be, e.g., interrupt enable writes (\texttt{ILE}/\texttt{IE} toggles), or a write to the \texttt{ENABLE} register of the peripheral.

\section{Evaluation}
This section presents an evaluation of our framework guided by three research questions. First, we assess the applicability of our tool to complex embedded workloads (RQ1). Second, we investigate its effectiveness in uncovering typical software bugs (RQ2). Lastly, we compare its performance against stateless \ac{vp} fuzzing to understand its advantages (RQ3).

\subsection{Experimental Results} \label{experimental_results}

\begin{figure}[t!]
\centerline{\includegraphics[totalheight=4cm]{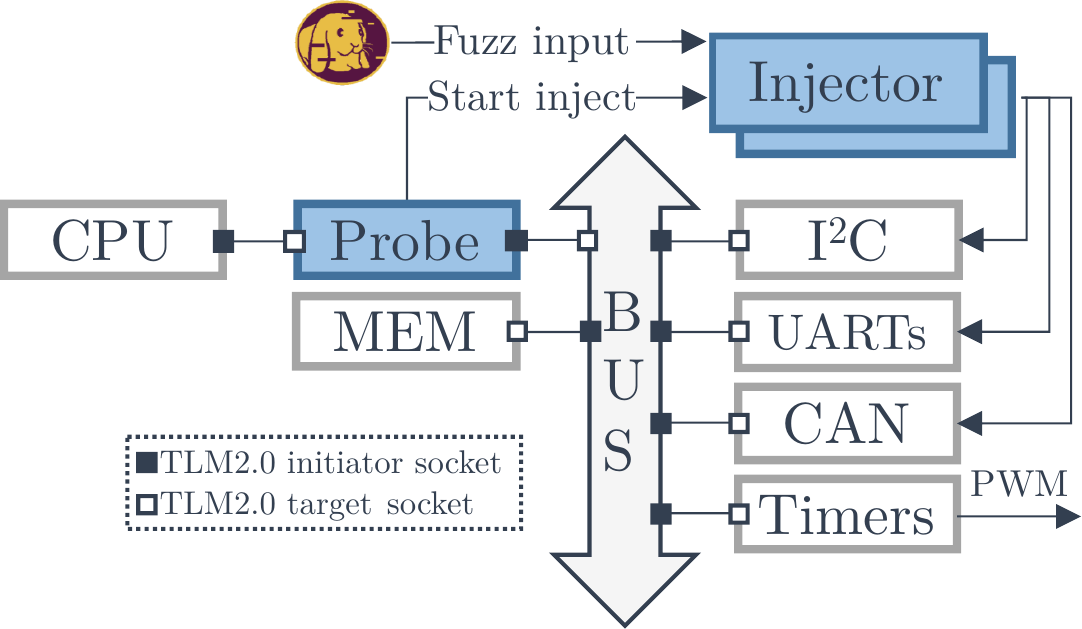}}
    \caption{The VP configuration for the four experiments.}
    \label{fig:baremetal}
\end{figure}

In our experiments, we tested unmodified and uninstrumented workloads as fuzzing targets exercising various peripherals and communication protocols. All experiments were conducted on a workstation with an Intel Core i7-1255U processor (4.7~GHz). Each target was executed on the SIM-A \ac{vp} using models of Nordic Semiconductor SoCs peripherals. SIM-A supports multiple ISAs, including Cortex-M and Cortex-A. For this work, we used the Cortex-M0 core. The models included in the simulator are those provided by the VCML library. Therefore, no additional modelling effort was necessary. The VP's modules and their connections are shown in Fig.~\ref{fig:baremetal}. No specific fuzzing seeds were given, each run began with a random byte sequence.

\begin{table}[b]
\centering
\caption{Conducted experiments.}
\setlength{\tabcolsep}{4pt}
\begin{tabular}{|llll|}
\hline
\textbf{\#) SW} & \textbf{Type} & \textbf{Peripherals}                                         & \textbf{Injected bugs}                                           \\ \hline
\rowcolor[HTML]{EFEFEF} 
A) Drone        & Bare metal    & \begin{tabular}[c]{@{}l@{}}I\textsuperscript{2}C, Timers,\\ UART\end{tabular} & \begin{tabular}[c]{@{}l@{}}Out-of-bounds\\ access\end{tabular}   \\
B) Robot        & Bare metal    & \begin{tabular}[c]{@{}l@{}}I\textsuperscript{2}C, Timers,\\ UART\end{tabular} & \begin{tabular}[c]{@{}l@{}}Out-of-bounds\\ access\end{tabular}   \\
\rowcolor[HTML]{EFEFEF} 
C) Passthrough  & Zephyr app    & UARTs                                                        & \begin{tabular}[c]{@{}l@{}}Invalid memory\\ access\end{tabular}  \\
D) Babbling     & Zephyr app    & CAN, Timer                                                   & \begin{tabular}[c]{@{}l@{}}Divide-by-zero\\ fault\end{tabular} \\ \hline
\end{tabular}
\label{table:1}
\end{table}

Table~\ref{table:1} summarizes the experimental setup. To evaluate RQ1, we chose as experiments the bare-metal firmwares of a drone and a robot~\cite{p2im}, A and B in Table~\ref{table:1} respectively, and two Zephyr \ac{os}~\cite{zephyr} applications, passthrough~\cite{passthrough} and babbling~\cite{babbling}, C and D. To address RQ2, we introduced typical embedded software vulnerabilities, such as out-of-bounds reads and invalid memory accesses, and evaluated whether our setup can effectively detect them. In addition to these injected bugs, our fuzzing campaigns also revealed two unknown bugs in the robot, drone and babbling targets.

\subsubsection*{Bare-metal Firmwares}
Firmware~A corresponds to the control software of a six-motor drone. It uses three timers that generate PWM signals for the six motors, one UART interface used for telemetry via an external Wi-Fi module, and an I\textsuperscript{2}C bus connected to an Inertial Measurement Unit (IMU), a barometer, and a magnetometer.
Firmware~B controls a two-motor self-balancing robot and follows a similar configuration: one timer driving the two motors, an I\textsuperscript{2}C bus connected to an IMU, and a UART linked to a serial console. Both firmwares were adapted to work with the \texttt{nrfx} driver stack~\cite{nrfx}.

For both firmwares, we implemented two input injectors. The first injector is attached to the I\textsuperscript{2}C peripheral and provides data when the firmware issues a read request. The second injector is associated with the UART peripheral and begins transmitting input as soon as the probe component detects an access to the UART \texttt{ENABLE} register. The timer peripherals only produce PWM output and do not consume external input. Therefore, no injectors are connected to them.

To evaluate our framework’s ability to detect software bugs (RQ2), we injected an out-of-bounds read vulnerability into the firmwares. Specifically, the code reads a sensor-reported data length without validating that the destination buffer has sufficient capacity, as shown in Listing~\ref{lst:out_of_bound}. In the simulated environment, no sensor was modeled, the fuzzer generated the \textit{len} value, which was then provided by the I\textsuperscript{2}C injector.

\begin{lstlisting}[language=C, caption=I\textsuperscript{2}C Out-of-bounds read., label={lst:out_of_bound}, basicstyle=\ttfamily\footnotesize]
void I2C_ReadBytes(uint8_t addr, uint8_t *data){
    // Read the length from the sensor
    uint8_t len = I2C_ReadByte(MPU6050, ACCEL);
    // Start the RX transaction to receive len bytes
    nrfx_twi_xfer(data, len);
}

\end{lstlisting}
This bug models a realistic class of issues in embedded drivers. Many real-world device drivers obtain a device-reported count or length and then perform multi-byte reads or allocations. For example, IMU sensors expose a register that reports how many bytes are currently in the device FIFO. The firmware reads this count and then issues a multi-byte read. Trusting that value without checking can cause an overflow of the host buffer. The injected out-of-bounds read produced an alignment fault on the VP, resulting from attempts to access an unaligned memory address beyond the buffer boundary.

During fuzzing of the firmwares, our tool repeatedly encountered unexpected cases where the system stalled after initializing the timer peripheral. The fuzzer observed that certain inputs caused the firmware to stop progressing, with no further timer activity. 
Detailed inspection revealed that the timer \ac{isr} was never invoked. Root-cause analysis traced this to a missing compile-time macro definition (\texttt{NRFX\_IRQ\_ENABLE}).
This defect caused a firmware hang, as it entered a busy-wait loop polling a flag that should have been set by the missing \ac{isr}. Both injected and unknown bugs were detected within the first 100 fuzzing executions. A summary of the injected and unknown vulnerability types, and their observed consequences is shown in Table~\ref{table:2}.

\begin{table}[b]
\centering
\caption{Description and effect of the bugs.}
\begin{tabular}{|l|ll|ll|}
\hline
                              & \multicolumn{2}{c|}{\textbf{Injected Bugs}}                                                                                  & \multicolumn{2}{c|}{\textbf{Pre-existing Bugs}}                            \\ \cline{2-5} 
\multirow{-2}{*}{\textbf{SW}} & Description                                                     & Effect                                                & Description                                                  & Effect \\ \hline
\rowcolor[HTML]{EFEFEF} 
A                             & \begin{tabular}[c]{@{}l@{}}Out-of-bounds \\ access\end{tabular} & \begin{tabular}[c]{@{}l@{}}Alignment \\ fault\end{tabular} & \begin{tabular}[c]{@{}l@{}}Undefined \\ macro\end{tabular}   & Hang    \\
B                             & \begin{tabular}[c]{@{}l@{}}Out-of-bounds \\ access\end{tabular} & \begin{tabular}[c]{@{}l@{}}Alignment \\ fault\end{tabular} & \begin{tabular}[c]{@{}l@{}}Undefined \\ macro\end{tabular}   & Hang    \\
\rowcolor[HTML]{EFEFEF} 
C                             & \begin{tabular}[c]{@{}l@{}}Invalid memory\\ access\end{tabular} & Bus fault                                                  & -                                                            & -           \\
D                             & Divide-by-zero                                                  & \begin{tabular}[c]{@{}l@{}}ARM hard \\ fault\end{tabular}  & \begin{tabular}[c]{@{}l@{}}Wrong frame \\ usage\end{tabular} & Hang    \\ \hline
\end{tabular}
\label{table:2}
\end{table}

\subsubsection*{Zephyr Applications}
We evaluated our framework on two additional targets that exercise the CAN and UART protocol stacks of the Zephyr OS (experiments~C and~D in Table~\ref{table:1}). These targets were selected because both protocols expose potentially vulnerable surfaces that receive external inputs, from other network nodes in the case of CAN, and from external devices in the case of UART.
In target~C, we implemented a UART passthrough application that bridges the console UART with a secondary UART connected to an external device. When one UART receives data, a callback writes the data into the other UART’s FIFO buffer. Two injectors emulate external devices, alternately sending and receiving data. The application uses the incoming UART data to compute an \ac{mmio} address (Listing~\ref{lst:invalid_access}), following a common pattern in drivers for multi-channel or buffer-based peripherals. The offset to the \texttt{MMIO\_BASE} address dynamically selects the target register. Without proper bounds checking, this behavior can lead to invalid memory accesses. The vulnerability detection using our framework took around 60 thousand executions (around ten minutes).

\begin{lstlisting}[language=C, caption=Invalid memory access., label={lst:invalid_access}, basicstyle=\ttfamily\footnotesize]
static void uart_receive(uint8_t* buf, uint8_t len){
    // Read len byte from the external device
    uart_read(buf, len);
    // Calculate the address based on the read byte
    uintptr_t addr = MMIO_BASE + buf[0];
    // Write to the potentially faulty address
    uint32_t *p = (uint32_t *)addr;
    *p = 0xDEADBEEF;  
}
\end{lstlisting}
In target~D, we implemented a CAN consumer application that processes incoming frames via a kernel message queue. When it receives a frame, the consumer extracts a ratio value, which it uses to schedule a periodic timer (Listing~\ref{lst:divide_by_zero}). This is a common scenario when using the CAN peripheral: \acp{ecu} often send a ratio asking another device to increase or decrease a sensor sampling rate. The CAN controller receives frames from an injector that periodically sends frames after the probe component in the \ac{vp} detects an access to the CAN Interrupt Enable (\texttt{IE}) register. Because the application does not validate the denominator of the ratio, injection of the byte value \texttt{0x00} triggers a divide-by-zero Usage Fault on the target CPU. This benchmark reproduces a common embedded-system anti-pattern: trusting device-provided data without defensive validation.
Additionally, fuzzing revealed a previously unknown defect in the CAN peripheral driver. When the fuzzer transmits a standard frame (with an 11-bit identifier) containing a payload longer than eight bytes, the driver erroneously drops the entire frame instead of truncating it. As a result, the application hangs while waiting for data that are never delivered. According to the Bosch CAN User Manual~\cite{mcan}, the correct behavior should be to accept the frame and silently discard any bytes beyond the eighth.

\begin{lstlisting}[language=C, caption=Uncontrolled use of frames., label={lst:divide_by_zero}, basicstyle=\ttfamily\footnotesize]
static void process_frame(struct can_frame *frame){
    uint8_t hz = frame->data[0];
    // Convert to ms
    uint32_t period_ms = (1000 / hz);
    // Schedule a periodic timer
    timer_start(&sample_timer, K_MSEC(period_ms));  
}
\end{lstlisting}

\begin{figure*}[t]
\centerline{\includegraphics[width=.9\textwidth]{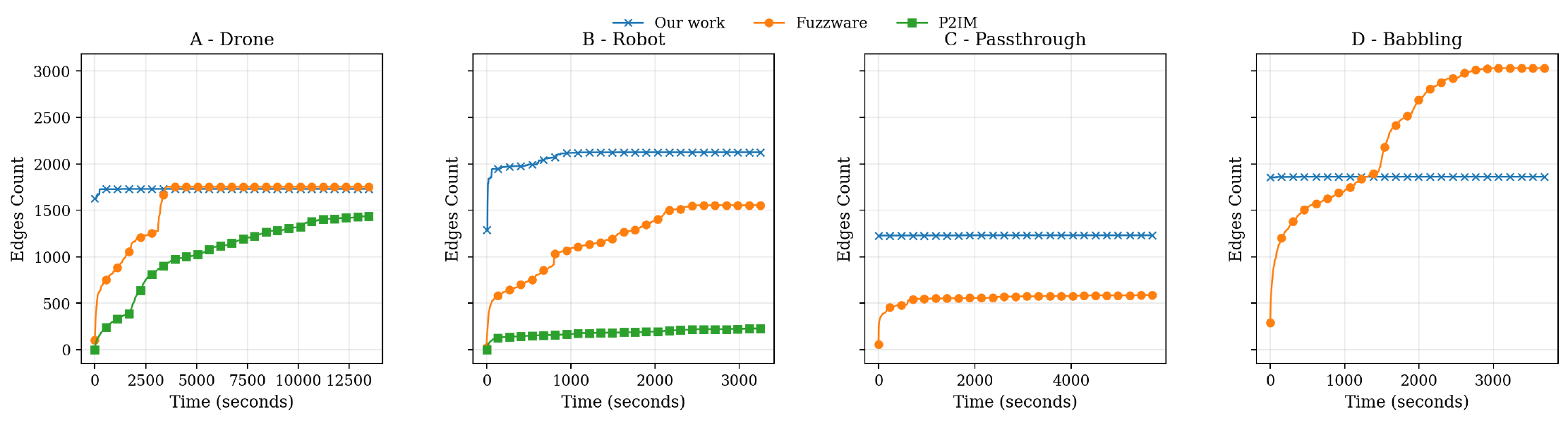}}
    \caption{Code coverage comparison for the various tools.}
    \label{fig:edge_comparison}
\end{figure*}

\subsection{Comparison with the State-of-the-Art} \label{comparison}
To evaluate RQ3, we compare our framework against two recent stateless \ac{vp}-based fuzzing tools, Fuzzware~\cite{scharnowski2022fuzzware} and P2IM~\cite{feng2020p2im}.
These tools were selected because they are open-source, published in top-tier venues, and enable a direct comparison between stateful and stateless approaches to \ac{vp} fuzzing (Section~\ref{related_work}). Like our framework, both aim to provide fuzzing inputs to embedded software targets to determine whether malformed inputs can drive the system into corner cases that reveal software defects.

\subsubsection*{Experimental Setup}
The metrics we consider are execution speed, number of false positives and code coverage. While the count of false positives can only be obtained via a manual reverse-engineering work, the code coverage and speed are provided by the AFL++ fuzzer.
Each tool was tested on the fuzzing targets listed in Table~\ref{table:1}, where the drone and the robot firmwares had already been tested by both tools in their works. Each fuzzing campaign was run until code coverage stabilized (no new edges discovered).
Table~\ref{table:3} summarizes the stabilized edge count and unique crash results. For comparability, we removed all artificially injected bugs and kept only the pre-existing bugs. Random seeds were used in all runs. P2IM could not be evaluated on targets~C and~D, as its firmware-analysis phase encountered state explosion and failed before fuzzing could begin.

\begin{table}[b]
\caption{Coverage and unique crashes comparison.}
\centering
\begin{tabular}{|l|llll|llll|}
\hline
                            & \multicolumn{4}{c|}{Edges count} & \multicolumn{4}{c|}{Unique crashes} \\ \cline{2-9} 
\multirow{-2}{*}{Framework} & A     & B     & C     & D     & A      & B       & C      & D       \\ \hline
\rowcolor[HTML]{EFEFEF} 
Fuzzware~\cite{scharnowski2022fuzzware} & 1750   & 1500   & 600   & 3100   & 1      & 14      & 0      & 876     \\ 
P2IM~\cite{feng2020p2im} & 1500    & 300    & -     & -     & 1      & 4       & -      & -       \\ 
\rowcolor[HTML]{EFEFEF}
Our work                       & 1750    & 2100    & 1250    & 1800    & 1      & 1       & 0      & 1     \\ \hline
\end{tabular}
\label{table:3}
\end{table}

\subsubsection*{Crash Inspection and False Positives}
As shown in Table~\ref{table:3}, Fuzzware and P2IM reported significantly more crashes than our framework, particularly in experiments~B and~D.
We manually analyzed a subset of the crashes and found that all inspected cases were false positives.
In experiment~B, P2IM reported four hangs. Investigation revealed that the firmware stalled in the UART transmit function, waiting for the Transmit Data Register Empty (TXE) flag to be set by the hardware. Because the simulation lacks a real UART model, the TXE flag is never set, causing the firmware to hang indefinitely.
The large number of false positives observed in Fuzzware can be attributed to two root causes:
\begin{itemize}
    \item Incorrect symbolic return values, e.g., receiving CAN frames that should have been filtered due to an ID mismatch, or erroneously returning 1 for \textit{exti\_is\_pending(line)}, leading to non-existent callbacks on the interrupt lines.
    \item Uncorrelated interrupt triggering, where interrupts are fired without the necessary preconditions, leading to failed \ac{isr} executions.
\end{itemize}
As shown in Table~\ref{table:3}, our tool eliminates false positives entirely, thanks to the peripheral models integrated into the full-system simulation. We were unable to evaluate the number of false negatives produced by our framework due to the large number of crashes identified by Fuzzware, of which we could analyze only a subset.

\begin{figure}[t]
    \centering
    \begin{tikzpicture}
    \begin{axis}[
        ybar,
        bar width=0.25cm,
        width=8cm,
        height=4cm,
        ylabel={execs/second},
        ymax=155,
        ymin=0,
        symbolic x coords={A,B,C,D},
        xtick=data,
        enlarge x limits=0.25,
        ymajorgrids=true,
        ylabel near ticks,
        legend style={
            at={(0.5,1.05)},
            anchor=south,
            legend columns=3
        },
        y tick label style={font=\small},
        y label style={font=\small},
        nodes near coords,
        every node near coord/.append style={
            anchor=west,
            rotate=90,
            font=\footnotesize
        },
    ]

    \addplot[draw=mplblue, 
             pattern=north west lines, 
             pattern color=mplblue,  
             line width=0.8pt,  
             bar shift=-0.35cm]     
             coordinates {(A,48)  (B,50) (C,95) (D,37)};
    \addplot[draw=mplorange, 
             pattern=north east lines, 
             pattern color=mplorange, 
             line width=0.8pt,
             bar shift=0cm] 
             coordinates {(A,101) (B,98) (C,120) (D,104)};
    \addplot[draw=mplgreen, 
             pattern=crosshatch, 
             pattern color=mplgreen,  
             line width=0.8pt, 
             bar shift=0.35cm]  
             coordinates {(A,47) (B,49) (C,0) (D,0)};

    \legend{Our work, Fuzzware, P2IM}
    \end{axis}
    \end{tikzpicture}
    \caption{\label{fig:speed}Comparison of the execution/second of the three tools.}
\end{figure}
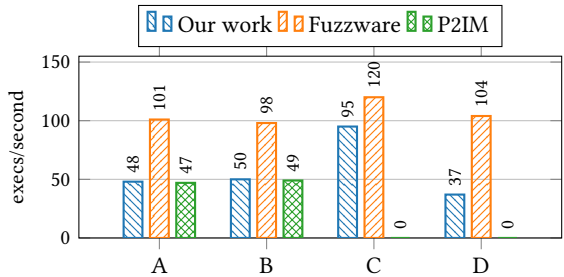
Figure~\ref{fig:edge_comparison} compares code coverage across the three tools.
Our framework achieves coverage comparable to Fuzzware for targets~A and~B, while Fuzzware reports approximately 40\% higher coverage in target~D. However, the inflated coverage is attributed to unrealistic execution paths, as indicated by the large number of false positives and spurious crashes on complex peripherals such as CAN.
In contrast, for target~C, our framework achieves 60\% higher coverage than Fuzzware. The reason is that Fuzzware lacks a UART model and therefore cannot emulate data transfers between UART instances, limiting its ability to reach deeper execution paths. The realism obtained from the inclusion of peripheral models comes at a cost: our framework is approximately 2× slower than Fuzzware, which uses a user-mode rather than full-system simulation. 
A comprehensive speed comparison is shown in Figure~\ref{fig:speed}.

\section{Conclusion}
Fuzzing has become one of the most effective techniques for discovering software vulnerabilities, thanks to its ability to automatically generate and test large numbers of inputs with minimal human effort.
In this work, we presented a framework that bridges fuzzing with full-system simulation to enable effective pre-silicon testing of embedded firmwares. Unlike existing stateless approaches that rely on symbolic or inferred peripheral models, our framework employs causally accurate peripheral simulation, enabling firmware execution under realistic hardware conditions without requiring source-code modification or instrumentation.

We evaluated our framework on a diverse set of realistic embedded workloads, including motor-control and sensor-driven applications, as well as components of the Zephyr OS protocol stack. Through controlled bug-injection experiments, we demonstrated that our approach reliably detects memory safety violations such as out-of-bounds reads and divide-by-zero faults. In addition, our framework successfully uncovered previously unknown bugs in unmodified firmware, highlighting its practical applicability for pre-silicon validation.

Compared to state-of-the-art stateless fuzzing tools, our approach achieves comparable code coverage while eliminating false positives, owing to its high-fidelity peripheral models. Although the full-system simulation incurs approximately 2× slower execution speed, the increased realism and diagnostic accuracy make it significantly more reliable for embedded testing. 

Future work includes extending the framework with automatic peripheral modeling starting from the vendor specifications, to easily expand the range of supported drivers, and enabling parallel fuzzing instances to improve performance.

\section*{Acknowledgements}
This work has been supported by funding from the Agentur für Innovation in der Cybersicherheit GmbH (Cyberagentur) under the 'Ecosystem Formally Verifiable IT – Provable Cybersecurity' (EVIT) Program (CAEU-WD/2023-63).


\clearpage
\balance
\printbibliography

\end{document}